\documentclass[preprint2]{aastex}
 
\begin{document}
\title{Interpreting the Behavior of Time Resolved Gamma-Ray Burst Spectra.}
\author{Nicole M. Lloyd-Ronning\altaffilmark{1} \& Vah{\'e} Petrosian\altaffilmark{2}}
\affil{Center for Space Science and Astrophysics, Stanford University, Stanford, CA
94305}
\email{nicole@urania.stanford.edu,vahe@astronomy.stanford.edu}

\altaffiltext{1}{Present Address: Canadian Institute for Theoretical
Astrophysics}
\altaffiltext{2}{Also Departments of Physics and Applied Physics}

\begin{abstract}
 In this paper, we explore time resolved
     Gamma-Ray Burst (GRB) spectra in the context of the synchrotron emission
     model presented in Lloyd and Petrosian (2000; LP00).  
    First, we show that 
     our model - which involves three distinct emission regimes -
     can provide excellent fits to the time resolved spectra
     of GRBs, and we present these results for a few bursts.  
     We then describe how the phenomenological Band spectrum (Band et al., 1993)
     can be interpreted in the context of our models based on the value
     of the low energy photon index $\alpha$.  We discuss the types
     of correlations one would expect to observe among the Band parameters
     if these models are correct.  We then compare these predictions to
     the existing data, 
    combining a sample of 2,026 time
     resolved spectra (from approximately 80 bursts).
     We show that the correlations found in the data are consistent with the 
     models, and discuss the constraints they place on the emission physics.
     In particular,
     {\em we find a ($\sim 4\sigma$) negative correlation between the peak
     of the $\nu F_{\nu}$ spectrum, $E_{p}$, and the low
     energy photon index $\alpha$} for bursts with
     $-2/3 < \alpha < 0$, in contrast to what is predicted
     by the instrumental effect discussed in LP00.
     We suggest that this correlation is simply due to the mechanism
     responsible for producing $\alpha$'s above the 
      value of $-2/3$ - namely, a decreasing mean pitch angle
     of the electrons.  We also show that  
      $E_{p}$ is correlated with the
     photon flux, and interpret this as a result of
     changing magnetic field or characteristic electron energy between
     emission episodes.
     Finally, we discuss the implications our results have on
     particle acceleration in GRBs, and prospects for further testing
     these models with the anticipated data from HETE-2, Swift and GLAST.\\
\end{abstract}

    \section{Introduction}
   Except for a few isolated bursts (see, for example, Tavani, 1996, Brainerd, 1996),
   most of the analyses of the 
   {\em prompt} spectral data of Gamma-Ray Bursts 
   have employed phenomenological models - in most cases, using the so-called
   Band spectrum (Band et al., 1993).  This model is
   essentially a smoothly broken power with a low energy photon spectral
   index $\alpha$, a high energy photon index $\beta$, and a break
   energy $E_{p}$. It  has been shown that such a model describes most GRB spectra
   very well (e.g. Mallozzi et al., 1996, Lloyd \& Petrosian, 1999, 
   Preece et al., 1999).  There have been
   some attempts to explain or interpret some of the global properties
   of these spectral parameters in
   terms of a physical model 
   (e.g., Totani, 1999, Preece et al., 1998, Ghissilini et al., 2000), with
   inconclusive results.
    In Lloyd and Petrosian (2000; herafter LP00),
     we showed that realistic synchrotron models can qualitatively
    explain the global distributions of the time
    averaged Band spectral parameters.  Our model modified the usual simple
    picture of optically thin synchrotron emission from a power
    law distribution of electrons with a sharp low energy cutoff,
    by accounting for: 1) the possibility of a smooth cutoff to the low 
    energy electron
    distribution, 2) radiation from an anisotropic
    electron distribution with
    a small mean pitch angle, 3) synchrotron self-absorption, and
    4) the important instrumental effect in which the value
    of the fitted
    parameter $\alpha$ decreases as $E_{p}$ approaches 
    the lower edge of the BATSE
    window.  We have envisioned a realistic scenario in which particle
    acceleration and synchrotron losses occur continually and simultaneously
    behind an internal shock (which produces a single pulse
    in the GRB time profile) with the characteristic acceleration
    time shorter than the loss time,
    so that synchrotron loss effects are only
    evident in the particle distribution spectrum at energies much
    larger than what is relevant for our discussion here (at energies where the inequality
    is reversed and the loss time becomes shorter than the acceleration
    time).
     This model stands up well to the global distributions
    of GRB parameters.  In particular,
    it can accommodate the bursts
    with $\alpha$ above the so-called ``line of death'' value
     $\alpha=-2/3$ (Preece et al., 1998).
    However, the tests conducted so far have involved time averaged properties of the
    bursts.  Since each pulse is regarded as a separate emission
    episode (or internal shock) in our model, the averaging over many pulses may obscure 
    the real physics of each
    episode. It would, therefore, be
    useful to examine the
    {\em time resolved} spectral properties of GRBs, so that one may
    compare the values and
    the \underline{correlations} between the spectral parameters from pulse to pulse and
    perhaps within a pulse, in hopes of gaining insight on the
    evolution of physical parameters throughout a GRB.

    The temporal evolution of GRB spectral parameters has been
    studied by several authors 
    (e.g. Norris et al., 1986, Kargatis et al., 1994,
    Ford et al., 1995, Crider et al., 1997,
    and Preece et al., 1998, Ryde \& Svensson, 2000).
      These studies have attempted to look for global trends
    in the data - in particular, Norris et al. and
    Ford et al. report a hard-to-soft
    evolution trend, while Crider et al. report both a hard-to-soft
    and ``tracking'' trend (in which one or more spectral parameters track
    the time profile of the burst).  The temporal behavior has not
    been rigorously interpreted in terms of an emission scenario (although
    Crider et al. attempt to explain at least the hard-to-soft evolution
    as evidence of a Comptonized spectrum in which the medium is
    expanding, causing  $E_{p}$ to decrease with time).
    Furthermore, with the exception of Kargatis et al., 
    these studies have perhaps over-simplified the extremely varied
   and complex evolutionary behavior seen in most GRBs.  
    
    The purpose of this paper is to use our synchrotron emission models
    as diagnostics for interpreting the 
    time resolved GRB spectral parameters.  
    Using a physical model - rather than a phenomenological
    model such as the Band spectrum - to characterize the GRB spectra, allows us
    to gain insight into the evolution of actual physical parameters
    in the GRB plasmas.  For example, in the case of optically thin emission
    by an isotropic 
    distribution of electrons, $E_{p} \propto \gamma_{m}^{2}B_{\perp}$,
    where $\gamma_{m}$ is the characteristic cutoff or
    turnover energy of the electron
    distribution, and $B_{\perp}$ is the perpendicular component of 
    the magnetic field.  The evolution of $E_{p}$ reflects the evolution
    of $\gamma_{m}$ and/or $B_{\perp}$.  Clearly, just examining
    the evolution of this one spectral parameter will not lead to insights
    on the evolution of a single physical quantity, since the former ($E_{p}$) depends
    on more than one physical variable ($\gamma_{m}$, $B_{\perp}$).  
However, examining the evolution
    of this parameter and its {\em correlation} with other spectral
    parameters can help break some of the degeneracies and the temporal
    behavior of the plasma parameters can begin to be elucidated.  
    
    Furthermore, once we get a handle on these parameters and their
    evolution throughout the GRB event, we can begin to gain some
    insight into the fundamental problem of particle
    acceleration in GRBs.  The particle acceleration and loss (e.g. to
    to radiation and otherwise) mechanisms determine the emitting electron distribution.
    The physics behind the particle acceleration determines the value
    of the minimum and maximum electron Lorentz factors, the smoothness
    of the low energy cutoff to the electron distribution, the high energy
    power law index $p$ of the electron distribution, and is of course intimately
    related to the magnetic field.
    The presence of a significant
    number of bursts with steep high energy spectra - that
    is, with a high energy photon index $\beta = -(p+1)/2$ 
    less than
    about $-3$ - suggests that $p$ 
    can be significantly larger than the so-called
     ``universal index'' $p \approx 2.2$ due to first
    order Fermi acceleration in a relativistic shock (see, e.g., Kirk et al., 2000 and
    references therein); thus, a different type of
    particle acceleration mechanism may be at work.  
    Furthermore, the evidence for the presence of synchrotron radiation
    from electrons with small pitch angles (see LP00 and below)
    also suggests that the usual assumption of isotropicization of the
    particle spectrum on short timescales may not be warranted.
    It is clear, then, that a burst's spectrum can help guide us in a 
    detailed study of particle acceleration in internal shocks.  
    
     In this paper we will further develop and explain our synchrotron
     emission models, and then use them to examine the behavior of
     the time resolved GRB spectra in the context of an in internal
     shock scenario.  In \S 2,
    we review the characteristics of our synchrotron models, and show that
    these models provide good fits to the existing data. In \S 3, we 
    discuss the types of correlations expected
   among the spectral parameters, in the context of our models.
    In \S 4,we examine correlations present between spectral
    parameters from Band fits carried out by Preece et al. (1999)
    in a sample of 2,026
    time resolved spectra.
   We find a
    strong positive correlation between the total photon 
    flux,$f_{\gamma}$, and the peak of the $\nu F_{\nu}$ spectrum, $E_{p}$,
    and correlations
    between $E_{p}$ and the low energy
    photon index, $\alpha$, which differ depending on whether $\alpha$ is above
    or below the value of $\sim -2/3$.
     We interpret
    these results, in the context of our models,
     as reflecting changes in the physical parameters from one emission episode
     to the next (one internal shock to the next).
     In \S 5, we
    present a few cases of spectral evolution for individual bursts, simply to
    illustrate how the models can be used to infer something about the
    physics in a particular burst (from shock episode to shock episode).
     Finally, in \S 6, we present conclusions
    and discuss the implications
    of our results on particle acceleration in GRBs.

 \section{Some Distinct Synchrotron Emission Scenarios}
   The details of our synchrotron emission models are described
   in LP00, where it is shown that the low energy spectral
   index $\alpha$ plays a key role in understanding the emission mechanism(s)
   at hand.   In this section, we describe three possible emission scenarios
   with distinct  
   asymptotic low energy spectral behavior.
   
\subsection{Isotropic Pitch Angle Distribution of Electrons (IPD)}
This is the familiar optically thin synchrotron emission 
from a 
power law  electron energy spectrum, with an isotropic pitch
angle distribution; but, in contrast to 
most analyses, here we consider an electron distribution 
 with a smooth low energy cutoff: $N(\gamma) \propto
\frac{(\gamma/\gamma_{m})^{q}}{1+(\gamma/\gamma_{m})^{p+q}}$. 
Note that for high energies ($\gamma > \gamma_{m}$),
 the spectrum goes as $\gamma^{-p}$, while
for low energies ($\gamma < \gamma_{m}$), the spectrum goes
as $\gamma^{q}$. Hence, $q$ denotes the 
steepness of the
electron low energy cutoff (note that an actual ``cutoff'', in the sense
that $N(\gamma) \rightarrow 0$ as
$\gamma \rightarrow 0$, requires $q > 0$).  The
asymptotic behavior of the synchrotron (photon number)
spectrum for $q > -1/3$ is:
\begin{equation}\label{equation} F_{\gamma} = \cases{\nu^{-2/3}&
$\nu \ll \nu_{m} = \frac{2}{3} \nu_{B} {\rm sin}\Psi \gamma_{m}^{2}$\cr
\nu^{-(p+1)/2}&  $\nu \gg \nu_{m}$}
\end{equation} 
where $F_{\gamma}$ is the photon flux, $\Psi$ is the electron pitch angle,
and $\nu_{B} = \frac{eB}{m_{e}c}$
where $B$ is the magnetic field.  Note that  the
peak of the $\nu F_{\nu}$ spectrum will occur at
$E_{p} \propto \nu_{m} \propto B{\rm sin}\Psi 
\gamma_{m}^{2}$, and that the aymptotic low energy index below this break 
is $\alpha = -2/3$.  We point out that if $q<-1/3$, the low energy asymptotic index
is $\alpha = (q-1)/2$; since $q<-1/3$ 
does not constitute a ``cutoff'', we do not discuss this further and
limit our discussion to cases with $0<q<\infty$.   However, even though the low energy
asymptotic index of the photon spectrum is always $-2/3$ for these (latter)
cases, this does not mean
that the value of $q$ does not play an important role in the observed
spectrum.  As shown in LP00 and discussed in
\S 3, the smaller the value of $q$ is, the lower the frequency
at which the asymptotic value of $-2/3$ is reached;
because of the finite width of the detector spectral
window, this can cause the fitted value of $\alpha$ to be significantly
less than $-2/3$.

\subsection{Small Pitch Angle
 Distribution of Electrons (SPD)} 
 This spectrum results from optically thin 
synchrotron emission by electrons with a mean pitch angle $\Psi \ll 1$;
the analysis of synchrotron radiation in this regime was first
done by Epstein (1973).
For high density, low magnetic field plasmas, the Alfv\'en phase 
velocity is less than the speed of light and (therefore) the speed of the 
relativistic particles under consideration here. In this case, the pitch angle diffusion 
rate of the electrons interacting with plasma turbulence is 
much larger than the acceleration rate; consequently, the accelerated electrons
will have an isotropic pitch angle distribution. However, for the {\em low
density, 
high magnetic field conditions} expected for the sources of GRBs, the opposite is
true. In this case the amplitude of the electric field fluctuations  
exceeds that of the magnetic field so that the above situation is 
reversed
(see e.g. Dung and Petrosian, 1994 and Pyradko and Petrosian, 1998). Then the 
pitch angle distribution of the accelerated electrons could become highly 
anisotropic as required in the small pitch angle model. The shape of
this spectrum depends on just how small the pitch angle is. For
$\Psi \ll 1$, but $\Psi\gamma_{m} \sim 1$, we have:
\begin{equation}\label{equation} F_{\gamma} = \cases{\nu^{0}&
$\nu \ll \nu_{s} = \frac{2}{3}\nu_{B}/(\gamma_{m}\Psi^{2})$\cr
\nu^{-2/3}&  $\nu_{s} \ll \nu \ll \nu_{m}$\cr
\nu^{-(p+1)/2}&  $\nu_{m} \gg \nu$}
\end{equation}
There are two breaks in this spectrum - one
at $\nu_{m}$ and one at $\nu_{s}$.  Because the Band spectrum can
only accommodate one break, spectral fits to
this model will put the parameter $E_{p}$ at one
or the other of these two breaks, but most likely at $\nu_{m}$ 
 because
for $p > 5/3$ (or for  high energy photon
index $\beta < -4/3$ which is the case for most bursts), the break across $\nu_{m}$ is more
pronounced than across $\nu_{s}$.  In this case, the low energy photon
index $\alpha$ will fall somewhere between $-2/3$ and $0$. 

 However, as
the pitch angle $\Psi$ decreases such that $\Psi \ll 1/\gamma_{m}$, then
the $\nu^{-2/3}$ portion of the spectrum disappears, and only the
$\nu^{0}$ portion is left.  In this case we have:
\begin{equation}\label{equation} F_{\gamma} = \cases{\nu^{0}&
$\nu \ll \nu_{s} =\frac{4}{3}\nu_{B}\gamma_{m}$\cr
\nu^{-(p+1)/2}&  $\nu_{s} \gg \nu$,}
\end{equation}
where $E_{p} \propto B\gamma_{m}$ (see Epstein, 1973 for a more
detailed description of the behavior of the spectrum in this regime).
[We note that Medvedev (2000) has developed a model in which 
the transverse deflections of electrons in highly non-uniform,
small scale magnetic fields are smaller than the electrons' relativistic
beaming angles ($\sim 1/\gamma_{e}$), so that the entire trajectory of
the electron is observable.  In this case, a so-called ``jitter'' spectrum
is obtained (Medvedev, 2000) which has some of the same low energy characteristics
as the SPD spectrum - in particular, the low energy photon index in this model
also has a value of $0$.]

\subsection{Self-Absorbed Spectrum (SAS)} 
If the magnetic field and density are such that the medium becomes
optically thick to the synchrotron photons with frequency $\nu < \nu_{a}$,
then, for $\nu_{a} < \nu_{m}$,
 we have the following spectrum:
\begin{equation}\label{equation} F_{\gamma} = \cases{\nu^{1}&
$\nu \ll \nu_{a}, $\cr
\nu^{-2/3}&  $\nu_{a} \ll \nu \ll \nu_{m}$,\cr
\nu^{-(p+1)/2}&  $\nu_{m} \gg \nu$}.
\end{equation}
In that case,  $E_{p} \propto \nu_{a} \sim 
10 (nl)^{3/5}B^{2/5}\gamma_{m}^{-8/5}\Gamma^{9/5}$ Hz,
 where $l$ and $n$ are the path length and particle density in the co-moving
 frame, and
 we have assumed an electron energy distribution index $p=2$.  [We have also
 assumed an isotropic distribution of electron pitch angles; for a small
 pitch angle distribution, the $\nu^{-2/3}$ portion in equation (4) would
 be replaced by  $\nu^{0}$.]  For $\nu_{a} >
 \nu_{m}$ we
just have one break at $\nu_{a}$ with a low energy photon index
of $\alpha = 3/2$ (in both the isotropic and small pitch angle cases).  The possibility of
self-absorption in GRBs is a controversial issue.
We have shown (LP00) that there 
are  bursts for which a self-absorbed spectrum is a better fit than an 
optically thin one.  We also found that in these cases, the absorption
frequency tends to be near the lower edge of the BATSE
window.  In addition to this,
Strohmeyer et al. (1998) found that a number
of bursts observed by GINGA with $E_{p}$'s in the
range 2 to 100 keV have steep ($\alpha \sim  1$) low energy spectral indices
consistent with a self-absorbed spectrum.
This raises interesting questions about the 
 physics of the ambient plasma, because self-absorption in a GRB
requires fairly large magnetic fields 
and particle densities.  For example, if the absorption frequency
is less than the minimum electron frequency, 
the optical depth to
 synchrotron self-absorption is
 \begin{equation}
  \tau \sim (l/10^{13}cm) (n/10^{8}cm^{-3})
 (B/10^{8}G)^{2/3}$$
 $$(\gamma_{m}/50)^{-8/3}
 (\Gamma/10^{3})^{3}(h\nu_{obs}/40keV)^{-5/3}(1+z)^{-5/3},
 \end{equation}
 where  $\nu_{obs}$ is the absorption frequency in the observer's frame, and
 $z$ is the redshift of the GRB.
  Note
 that
 this frequency falls within BATSE's spectral window under certain,
 perhaps somewhat extreme, conditions.
  The physical processes  required to achieve these conditions
 will need to be theoretically
 established if the data prove self-absorption 
 to be a viable model.  
  We point out that the next generation of
 GRB dedicated telescopes - namely SWIFT and HETE-2 - will obtain more
 spectral data in energy ranges lower than the BATSE threshold (of about
 25 keV) and can firmly establish the presence or absence of a self-absorbed
 portion of the low energy spectrum.
 
 Figure 1 shows the various spectra for the different emission regimes.
 Note that we have plotted the $F_{\nu}$ spectrum rather than 
 $F_{\gamma} = F_{\nu}/\nu$ to emphasize the differences in the various
 low energy slopes of the spectra between the different emission regimes.
 Throughout the rest of the paper, we use the value of the low energy photon
 index $\alpha$ to distinguish between the different emission scenarios,
 where the {\bf IPD case is
 defined by $\alpha \la -2/3$, SPD by $-2/3 \la \alpha \la 0$,
 and SAS by $\alpha \ga 0$}.
 
 \begin{figure}[h]
\plotone{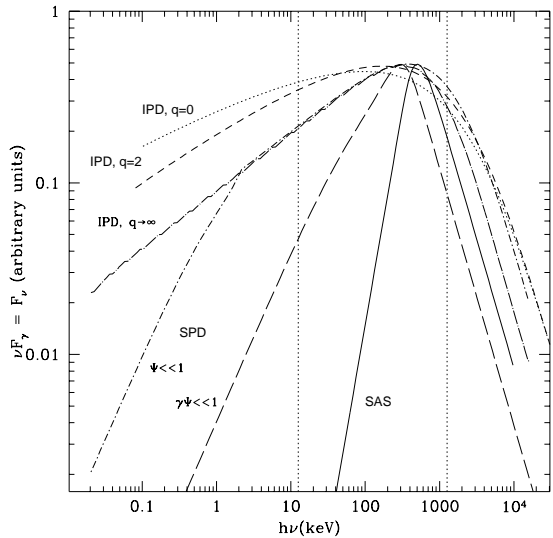}
 \caption{  Various synchrotron energy spectra, $F_{\nu}$ (in arbitrary
units), as a function
of energy $h\nu$ in keV. The dot-dashed line is 
  optically thin radiation from an isotropic
  distribution of electrons (IPD) with a sharp minimum energy cutoff,
   while the dotted and short-dashed
  lines show the IPD synchrotron spectra for smooth cutoffs to
  the electron distribution ($q=0$ and $q=2$  
  respectively).  Note that for finite values of $q$,
  the asymptotic $\nu^{1/3}$ spectrum is achieved at photon
  energies much lower than $E_{p}$.  The solid line shows a 
 self-absorbed spectrum (SAS)
 for $\nu_{a}>\nu_{m}$. The long-dashed and dot-dashed
 lines indicate the
 small pitch angle distribution (SPD)
  case for small ($\Psi \ll 1$) and very small ($\gamma\Psi \ll 1$)
  pitch angles, respectively.  
The vertical lines mark the approximate width of the BATSE spectral window.}
 \end{figure}
 
 \subsection{Spectral Fits - Directly Testing the Model}
 
 To test the 
 how well these synchrotron models actually describe the existing data,
 we have performed time resolved spectral fits to a sample of data from the 
 BATSE archive.  We use 128 channel, 128ms time resolved HER data, which is
 obtained for the most brightly illuminated
 of the eight detectors from the on-line archive at: 
 cossc.gsfc.nasa.gov/compton
 /data/batse/trigger. 
  We plot up the total counts as a function of time (the
 burst time profile) summing over all energy bins; from this, we pick out time
 intervals for the background and over which to do our spectral fits.
We then subtract off the background counts
 averaged over our specified ``background'' time
 intervals from our raw spectral counts data for each energy bin
 (the alternative
 method of subtracting a fitted background photon
  model off of the spectral photon model is mathematically equivalent -
  that is, it is equivalent to
 subtract off the background before convolving with the detector
response matrix (DRM) 
or after convolving with the DRM ).
 We then fit our photon
 models to the data by convolving them with the DRM to get model counts and
 then minimized $\chi^{2} =$ (data-model counts)$^{2}/\sigma^{2}$,
 via a downhill simplex method.
 
 \begin{figure}[t] 
\plotone{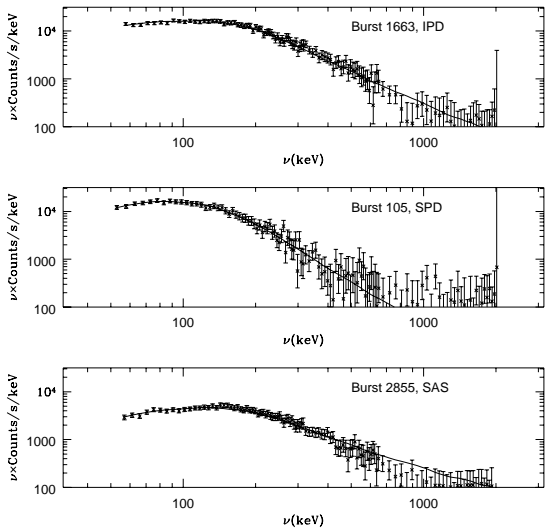}
 \caption{Fits to the three different synchrotron models described in \S 2.
The emission regime that best fits each burst (indicated in each figure) is
the regime inferred from the Band fit value
of $\alpha$ at the time of each spectrum.  [Note
that these are the actual counts spectra of the burst, which is the photon
spectrum convolved with the detector response matrix.]  }
 \end{figure}
 
 Our models are described in detail in LP00 and briefly above.
   We fit each spectrum to all 3 emission
 scenarios and then evaluate the fits based on their values of
 a reduced $\chi^{2}$. 
  In Figure 2, we show examples of spectral fits in each emission regime.
 Each fit is taken at a time during the burst spectral evolution when
 the $\alpha$ parameter corresponded to the respective model.  For example,
 in the top panel - burst 1663 - the spectrum is from a time when $\alpha \sim
 -2/3$,
 while in the middle panel - burst 105 - the spectrum is from a time in
 the profile when $\alpha = 0$.  Similarly, for the bottom panel, this spectrum
 corresponds to a time when $\alpha=1$.  The reduced $\chi^{2}$ are $0.34$,
 $0.33$, and $0.50$ for the top, middle and bottom panels respectively.
  In general the best model turns
 out to correspond to
 the emission regime suggested by Band's $\alpha$ values, which confirms
 our proposed method of  physically interpretating  Band fits based on
 the bursts' low energy photon index  
 (for example, an IPD fit to the spectrum of burst 105
 gave a $\chi^{2} > 1$ compared to the $\chi^{2} = 0.33$ for an SPD fit). However,
 we have not yet been able to confirm this for the self-absorbed case;
 although in some bursts the SAS was clearly better than the IPD case, SPD seemed to do statistically
 as well
 for the small sample of bursts we have tried.
As mentioned above, better low energy data is needed to definitively establish
the existence of a self-absorbed component in some GRB spectra. Nonetheless,
in this paper, we interpret
those bursts with a low energy photon index larger than the SPD limit
of $\alpha = 0$ as SAS cases.


 \section{Expected Correlations Among Spectral Parameters}
 Ideally, we would like to carry out such fits to a large sample of bursts,
 and characterize the physics of each burst as well as trends among large
 samples of bursts directly through the physical parameters
 yielded by the fits.  However, this is a large and time-consuming
 task, and in fact we can learn a great deal
 about the underlying physics of GRBs through an
 analysis of the phenomenological
 spectral parameters in terms of a physical model.  We have already shown
 that the Band parameter $\alpha$ is a good indication of the relevant
 synchrotron emission regime (IPD, SPD, or SAS).  Examining correlations
 among the various Band parameters can lead to additional insights on the
 physics governing the emission, when interpreted in the context of a physical model.
 Below, we discuss the types of correlations one might expect 
 among the Band spectral parameters for the different emission scenarios described above.
 In \S 4 and 5, we compare these correlations with what we find in the GRB data.
 
 \subsection{$\alpha-E_{p}$ Correlation}
 Two different correlations are expected for these parameters.
 
 {\bf 1.}
 We expect a positive 
 correlation between $E_{p}$ and $\alpha$ due to the instrumental
 effect described in LP00.  If $E_{p}$ is close to
 the edge of the BATSE window, the low energy photon index may not 
 yet have reached
 its asymptotic value and a smaller (or softer) value of $\alpha$
 (relative to the asymptotic value)
 will be determined.  A smooth cutoff to the electron energy
 distribution will exacerbate this effect because for a smoother
 cutoff (or a lower $q$), the low energy asymptote is reached
 farther away from $E_{p}$, nearer to (or even below) the low energy edge
 of the detector spectral window. 
  Note that a dispersion in the
 smoothness of the low energy cutoff will tend to wash this correlation
 out to some degree, as seen in Figure 4 of LP00. 
 For the cases of small pitch angle radiation and the self-absorbed
 spectrum, this effect will be weaker because the low
 energy asymptotes are reached more quickly (i.e. at energies closer
 to $E_{p}$) than
 for the isotropic optically thin case (see Figure 1).
 
 {\bf 2.} 
 We also expect evidence of 
 a negative correlation between $E_{p}$ and $\alpha$ as we transition
 from the IPD to the SPD regime, i.e.  for $-2/3<\alpha<0$.  In this case, the pitch angle  
 decreases so that $E_{p} \propto {\rm sin}\Psi$ decreases, if all other
 physical parameters ($B$ and $\gamma_{m}$) remain constant. In
 addition, as we
go from the small pitch angle regime, $\Psi \gamma_{m} \sim 1$ ($\Psi \ll 1$), to
the very small pitch angle regime,  
to $\Psi \gamma_{m}
\ll 1$, the $\nu^{-2/3}$ portion of the
spectrum disappears, and we are left with only the $\nu^{0}$ portion.
In other words, as the mean of 
the pitch angle distribution decreases to very small values,  $E_{p}$ decreases
and the value of $\alpha$ decreases from $-2/3$ to $0$.
   This negative correlation will compete with the
 positive instrumental correlation mentioned above.
 
 \subsection{$\beta-E_{p}$
 Correlation}
  We expect a similar correlation between $E_{p}$ and $\beta$ due
 to 
 instrumental effects. A
 dispersion in the high energy electron index $p$
  will tend to reduce this correlation.  However,
 in
 practice - partly because the high energy spectral data are not very
 constraining and partly because  $E_{p}$ is usually well below
 the upper edge of the BATSE window (about $1.5$ MeV) for those bursts
 with spectral fits - this correlation between $E_{p}$ and
 $\beta$ is not evident in
 the data.  For a few sample spectra, we find that $E_{p}$ has to
 be greater than around $1100$ keV before $\beta$ is affected by
 this instrumental effect. 

 \subsection{Total flux - $E_{p}$ Correlation}
 We might expect a positive correlation between $E_{p}$ and the
 flux of
 the burst.
 If $E_{p}$ changes either due to a change in
 the magnetic field or $\gamma_{m}$, then the flux, which also depends on
 positive powers of both of these parameters, 
  will also increase (Pacholczyk, 1970).
 This effect of course will be weakened to some degree by the
 distribution of redshifts of GRBs (if we examine the whole
 spectral sample instead of one  burst).  However, as
 shown in Lloyd et al., 2000, 
 the cosmological contribution to such a correlation (higher redshift
 reduces the observed value of $E_{p}$ and flux for a given burst)
 is negligible due to the large intrinsic dispersion in the luminosity
 function and intrinsic $E_{p}$ distribution,
  so that any correlation we do observe can be attributed to
 an intrinsic effect.

\section{The Time Resolved Spectral Data - Global Behavior}

  We showed in LP00 that the time averaged spectral parameter
  distributions are consistent with the models of synchrotron emission
  described above.  We also showed for several GRBs,
  that the evolution of different spectral
  parameters of a particular burst track eachother throughout their time evolution 
  in a way that is easily interpreted in our models.
 We want to test our models and in particular investigate the latter point
 in more detail by examining the behavior of a large sample of 
  {\em time resolved}
  spectral parameters.  
  The ideal is to learn something about how the plasma parameters are
  changing between emission episodes in a burst, by examining the time
  resolved of the Band spectral parameters in the context of our synchrotron
  models. As mentioned in \S 3, because the spectral parameters can depend on more
  than one physical quantity, we look for 
 {\em correlations between the spectral parameters} (rather than
examining the evolution of one spectral parameter in time),
  in order to break some of the degeneracies in interpreting
 the evolution of the physical parameters.

One way to do this is to look for particular trends between
pulses within  individual bursts. For example, Crider et al. (1997) 
claim to see a positive correlation
between $E_{p}$ and $\alpha$ in 47 individual bursts\footnote{We note that
they do not distinguish between different emission regimes in their
analysis.  They have also
found evidence for a negative correlation in some individual bursts, but with admittedly low
statistical significance.  They do not report the $\alpha$ values for  these bursts
(so that they might be interpreted in terms of a particular ``emission regime'' of
our models).}.
 The advantage of looking at spectral evolution within individual bursts
 is that correlations
  between the spectra do not contain
 any dispersions or contributions that might arise due to redshift effects.
 However, looking for global evolutionary trends
 in the data by examining individual bursts is a
 difficult and challenging task, particularly when we differentiate
 between different emission regimes.  This is not only
 because there are a small number
 (typically $\sim 20$) of time resolved spectra
 per burst, but also because 
 there are a small number of points {\em per emission regime},
 particularly for the SPD and SAS regimes. Therefore,
 attaching a significance to correlations between
 spectral parameters in different emission regimes
 for a single burst is in general not statistically robust (Efron, private communication)
 and there is therefore no reliable way to compare results with other bursts in order
 to establish general trends in the evolution.  
 
 We would like to investigate if there are any average trends present among the
 spectral parameters in each emission regime.
Although we do examine the behavior of some individual
bursts in the next section, this goal is best accomplished by 
 combining all 2,026
 time resolved spectra available, and searching for any global trends
 in this sample.  Trends present
 in this entire sample will
 reflect the trends of
 temporal behavior in individual bursts on average, and can therefore
 help us gain insight on the evolutionary trends present in
 individual bursts.  We can then test whether these trends are consistent
 with our models of synchrotron emission, and - if so - ideally
 learn something about the evolution of the physical conditions (such
 as the magnetic field and Lorentz factors) throughout a GRB.
 
 \subsection{Data and The $\alpha$ Distribution}
  Our data is taken from the catalog of Preece, et al. (1999), which
  contains high energy resolution,
  time resolved spectral fits to a large number of BATSE bursts (see their
  paper for discussion of data type, time and energy resolution, etc.).  Our sample
  consists of individual spectra for which the HER data type is
  used in the spectral fit (because of its superior energy resolution;
  see Preece et al., 2000 for description of data types), and
  for which the Band function provided a reliable
  fit in the $\sim 20$ keV to $\sim 1.5$ MeV range. These criteria leave
  us with $2,051$ spectra.  We then elimate 
  $25$ additional individual spectra, because the error bars on the
  spectral parameters are $0$, indicating an error in the fitting procedure.
  This leaves us with $2,026$ spectra; although this is only a fraction
  of the $\sim 5,000$ spectra in the Preece catalog, we believe it is
  an accurate representation of at least those bursts which are described
  by the Band spectrum.
  And as discussed in Preece et al. (1999), this spectral form
  describes the large majority of bursts very well (see also Band et al., 1993,
  Mallozzi et al., 1996, Lloyd and Petrosian, 1999)\footnote{ 
  For some bursts, however, this is not the case.  For example, Preece et
  al. (1996) showed that in a sample of about $90$ {\em time averaged}
  spectra, about $14\%$ showed an X-ray excess in the $7-20$ keV range.
  Strohmeyer et al. (1998) also found an X-ray excess in at least one
  GINGA burst.  Such an excess would tend to affect the low energy photon index - 
  giving a softer (lower) value of $\alpha$ than if there were no excess.
  This, in turn, would add some scatter to the expected correlations between
  $\alpha$ and $E_{p}$ discussed in $\S 3.1$.  However, because we
  are looking at the subset of data for which the Band spectrum provides a
  reliable fit, and we are examining {\em time resolved} spectra between
  $20 - 1500$ keV (above the range in which theses excesses have been found),
  we do not expect this would significantly affect our results.  Of course,
  one can always add additional emission components to any model; but
  because we have no evidence of it in our data, we take the simplest interpretation
  of a single component emission model and see what we can learn
  from the data under this assumption.}.
  
  \begin{figure}[t]
\epsscale{1.0}
\plotone{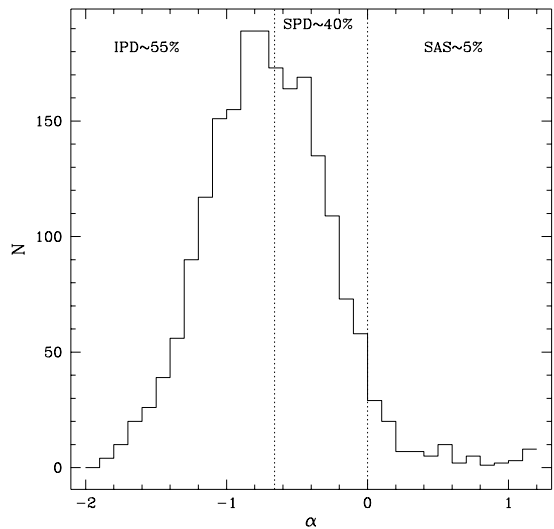}
\caption{Histogram of low energy photon index $\alpha$, a compilation
of 2,026 time resolved spectral fits from Preece et al., 1999.  The different
emission regimes and percent of spectra in each regime are marked on the
figure.}
\end{figure}

 \subsubsection{The $\alpha$ Distribution}
  As discussed in \S 2, the parameter
   $\alpha$ is the best parameter for distinguishing between
the various synchrotron regimes.
 Figure 3 shows a histogram of $\alpha$ (taken from the 2,026
  time resolved fits of Preece et al., 1999) with each regime 
  clearly marked\footnote{We point out that the peak of this distribution
  falls just slightly higher than the $\alpha$ distribution presented
  in Preece et al., 2000, which peaked at around $-1$.  This may be partly
  due to the differences in the way we binned our data, but primarily
  reflects the fact that we have only included those bursts with Band fits
  in our sample; for example, bursts that are described well by a ``Comptonized'' 
  spectrum or simple broken power-law spectra tend
  to have a slightly lower value of $\alpha$ than those described by the Band
  spectrum, and may shift the peak of the $\alpha$ distribution toward
  lower values. In any case, we note the difference is within the average $1\sigma$
  error on $\alpha$ ($<\Delta\alpha> \sim 0.28$), and remind the reader we
  are only examining those bursts for which the Band function provided an
  acceptable spectral fit.}. 
 For our sample, we find the percent of spectra in each regime is
  as shown in Table 1.
\begin{table}[h] 
\begin{center}
\begin{tabular} {lccr} \hline \hline
Regime & $\alpha$ range    & \#  & \%  \\ \hline
IPD  &  $\alpha \leq -2/3$  & 1122  & 55 \%  \\
SPD & $-2/3 < \alpha \leq 0$ & 805  &  40 \%   \\
SAS  &  $0 < \alpha < 3/2$  &  99 &   5 \%  
 \\ \hline \hline
\end{tabular}
\end{center}
\caption{Number spectra in each emission regime.}
\end{table}
Note that there are a significant number of spectra in the SPD regime 
 and SAS regime.  It is important, however, to
 briefly discuss how the error bars on the parameter $\alpha$  
 affect the interpretation of this distribution.  First, we point
 out that Preece et al. (1999) showed that the
 error bar on $\alpha$ cannot alone account for the large dispersion in the distribution.
In our sample, we find that the average $1\sigma$ error on $\alpha$ is
$<\Delta\alpha> \sim 0.28$, suggesting (in agreement with Preece et al.'s more
detailed analysis) that there is still a significant number 
of bursts above the line of death $\alpha=-2/3$.  To quantify this,
we have computed the number of bursts in each emission regime, using
$\alpha$'s upper and lower $1\sigma$ limits. If we take all values of $\alpha$ at their
upper limits ($\alpha+\Delta\alpha$), we find $42\%$, $46\%$, and $12\%$ of 
bursts in the IPD, SPD,
and SAS regimes respectively.  Taking all values of $\alpha$ at their
$1\sigma$ lower limits ($\alpha-\Delta\alpha$), we find $69\%$, $29\%$, and $2\%$ of 
bursts in the IPD, SPD,
and SAS regimes respectively. 
 Although the error bars on $\alpha$ can make some
difference as to the numbers of spectra in each regime, we will see
that this does not affect  the
qualitative nature of our conclusions below.
 We now discuss the correlations present in the data and their consistency with
 what we expect in
 the context of the three synchrotron emission scenarios.

 \subsection{Observed $\alpha-E_{p}$ Correlation}
 Figures 4a and b  show the binned average correlation between $E_{p}$ and $\alpha$
present in the time resolved data. For each of these figures, we have sorted $\alpha$
in ascending order and binned the data every 100 points (the horizontal
error bars indicate the size of the bins).  We then computed the
average and median $E_{p}$ for these 100 points.  In Figure 4a, we show the
average $E_{p}$ vs. $\alpha$, where the vertical
error bars are simply  the variance of the
mean value of $E_{p}$; in Figure 4b, we have plotted the median $E_{p}$,
where the solid and dotted vertical error bars indicate the range
of $E_{p}$ about the {\em median} value that includes $68\%$ 
and $90\%$ of the data respectively (we do point
out that the scatter of $E_{p}$ in each bin is not necessarily Gaussian; see, e.g.,
Preece, et al., 1996 and Preece, et al., 1998b).
The observed trends are 
 consistent with what is expected from our model in each
 emission scenario, and tell us something important about the role various effects
 play in the correlations, as we discuss below.
\begin{figure}[h]
\epsscale{0.71}
\plotone{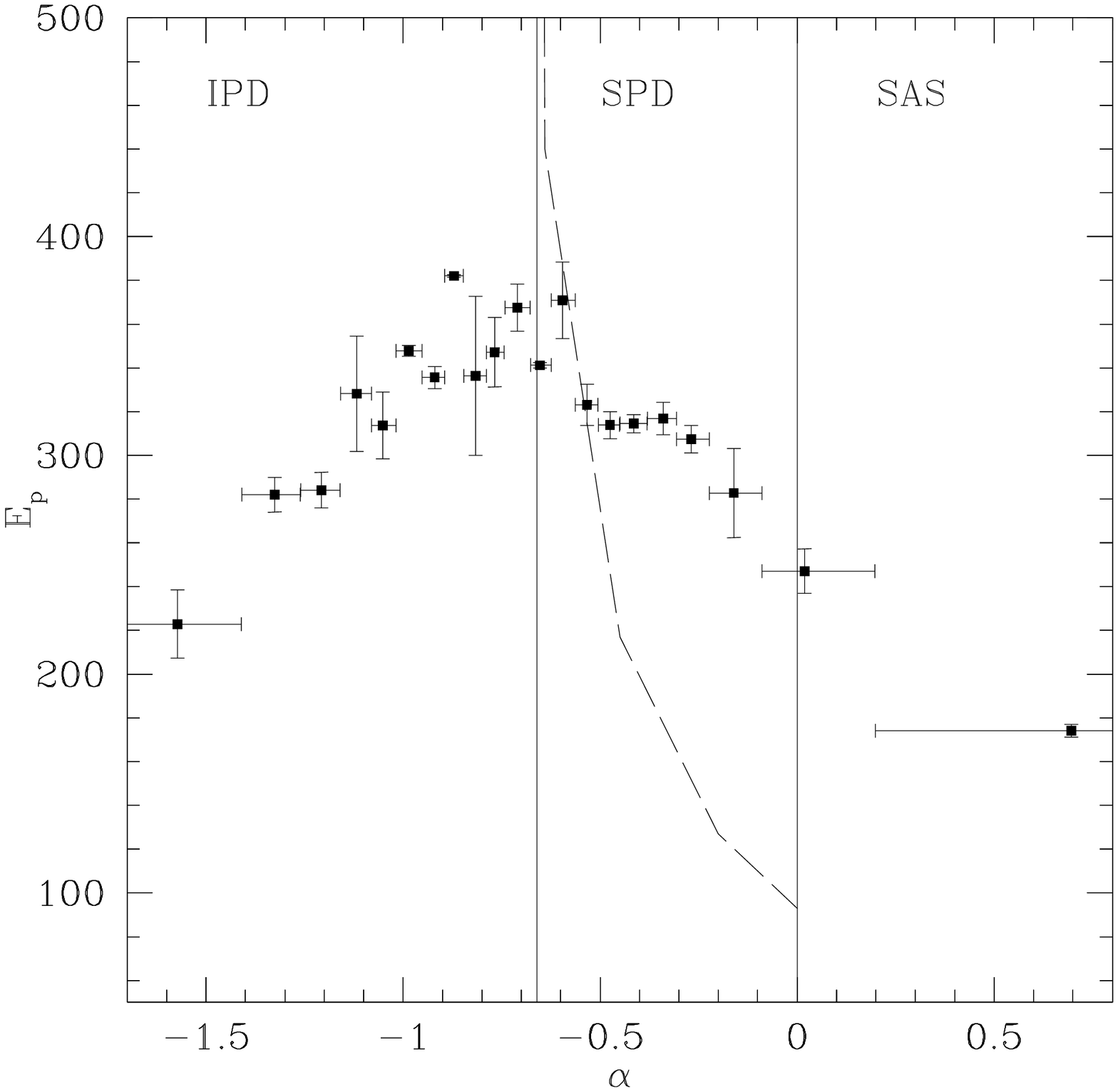}
\plotone{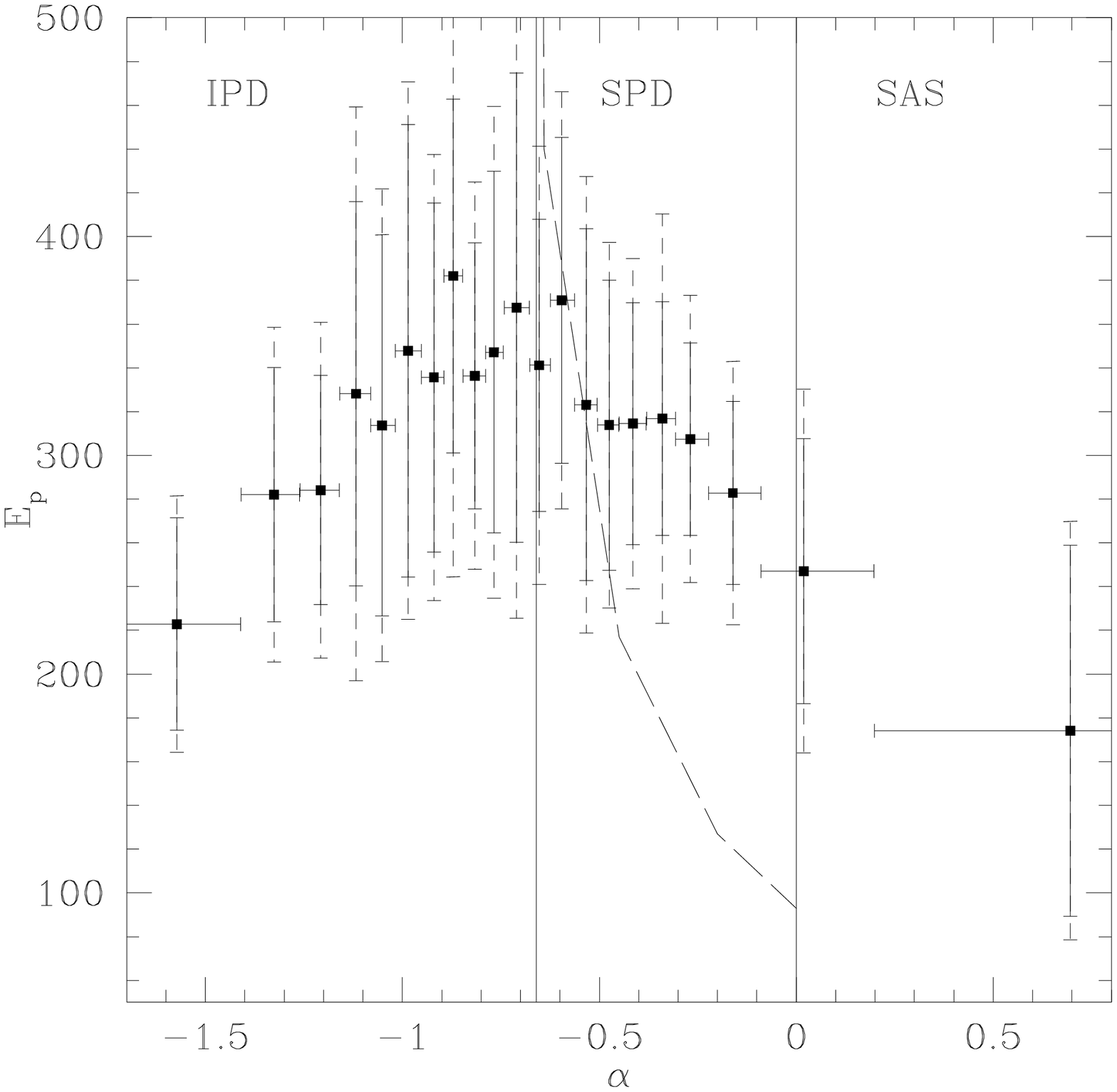}
\caption{ Peak of $\nu F_{\nu}$ spectrum $E_{p}$ vs. low energy index
$\alpha$ for binned 2,026 time resolved spectral fits.  The upper panel
plots the variance of the average $E_{p}$ as the vertical error bars,
while the lower panel plots the median values of $E_{p}$
and the range that contains $68\%$ (solid lines) and $90\%$ (dotted lines) of
the data around the median. Each regime of
emission is marked on the plot. The dashed line in the SPD regime is an example
of how $E_{p}$ changes as a function of $\alpha$ within the BATSE spectral
window as the mean of the electron pitch
angle decreases, but all other parameters ($B$, $\gamma_{m}$, $p$, $q$) remain constant.}
\end{figure}

\subsubsection{IPD Regime} 
Performing a Kendell's $\tau$ test
on all of the (unbinned) data, we find a $9 \sigma$ positive correlation
between $\alpha$ and $E_{p}$ in the IPD regime.  To account for both
the error in $\alpha$ and $E_{p}$, we have performed this test on
all permutations of correlations between the lower and upper values of $\alpha$ (from
the $1\sigma$ error bars) with the lower and upper values of $E_{p}$. In
addition we have averaged the value of $\tau$ from 
 100 sets of data, in which - for each data point - 
$\alpha$ and $E_{p}$ are drawn from Gaussian distributions
with means equal to the parameter
 values given in the catalog and standard deviations corresponding to
the error bars. for each of the data points, correlating these deviates, and taking
the average value for the correlation.  {\em In all cases, we find a highly
significant ($>6\sigma$) correlation}.  
The positive correlation between $\alpha$ and
$E_{p}$ in the IPD regime
can be simply understood by the instrumental effect
discussed in \S 3 above.
  However, for a given value of the low energy
 electron distribution cutoff parameter $q$, the
 correlation is expected to be much stronger than
 observed (see Figure 4 of LP00).  Of course,  
 any dispersion in $q$ will tend to weaken and flatten
this correlation. In fact, we can give a rough but
quantitative estimate of the $q$ distribution required to
produce the observed $E_{p}-\alpha$ correlation seen in Figure 4.  
 For each burst 
(under the assumption that optically thin synchrotron emission from
an isotropic distribution of electrons produces its spectrum),
we can estimate the $q$ value necessary to produce
the burst's $E_{p}$ and $\alpha$ values, simply from the determined relationships  
between $\alpha$ and $E_{p}$ for different values
of $q$ in Figure 4 of LP00.  
  In Figure 5, we present
an estimate of the distribution of $q$ required to reproduce the data and
therefore 
the shallow correlation between $\alpha$ and $E_{p}$ observed
in the IPD regime.  Note that the dotted part of the histogram is not at all well constrained
because the correlation dramatically weakens for very high values
of $q$ and the solutions become quite degenerate (that is, for
a given $E_{p}$, a $q$ of $5$ or $10$ may produce the same value
of $\alpha$). 
\begin{figure}[t]
\epsscale{1.}
\plotone{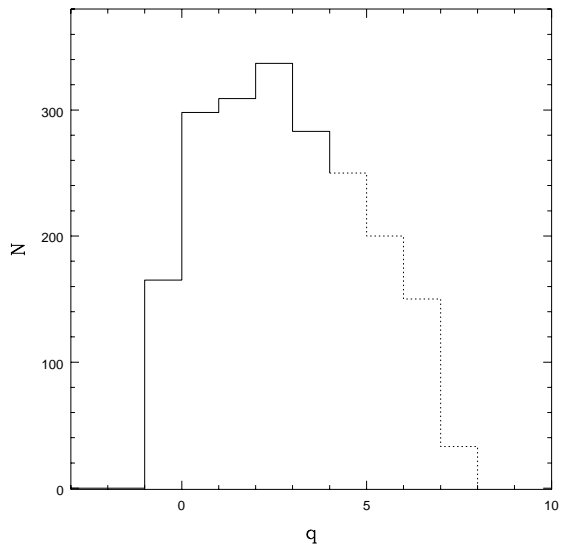}
\caption{Estimated distribution of $q$, given the observed correlation between
$\alpha$ and $E_{p}$ in the IPD regime. There is a degeneracy
among high values of $q$ (that is, for different $q$'s $ \ge 5$,
we obtain the same observed $\alpha$ value for a given $E_{p}$); the
dotted portion of the histogram attempts to account for this degeneracy
and is simply a reasonable guess at how this portion of the histogram behaves. }
\end{figure}

\subsubsection{SPD Regime}
The sign of the correlation between $\alpha$ and $E_{p}$  reverses
for values of $\alpha \ga -0.7$, which is suggestively very close
to where we expect a transition from the IPD to SPD regime.
Performing a Kendell's $\tau$ test
on all of the (unbinned) data in this regime, we find a $4 \sigma$ {\em negative} correlation
between $\alpha$ and $E_{p}$.  Again, to account for  
the error in both $\alpha$ and $E_{p}$, we have performed this test on
all permutations of correlations between the lower and upper values of $\alpha$ (from
the $1\sigma$ error bars) with the lower and upper values of $E_{p}$, as well
as averaged the $\tau$ value from 100 sets of data drawn from distributions
based on the existing data, according to the prescription described in $\S 4.2.1$. 
 {\em In all of these cases, we find a  
significant ($>3\sigma$) negative correlation}.
As mentioned in \S 3, this type of correlation is natural in the
small pitch angle regime, simply as a result of decreasing average pitch
angle.
As seen in equation (2) and described in \S 3.1, when $\Psi$ decreases,
the characteristic SPD frequency
$\nu_{s}$ approaches the characteristic IPD
frequency $\nu_{m}$ causing the fitted value of $\alpha$ to
increase from $-2/3$ to $0$. 
Meanwhile, $E_{p} \propto \nu_{m} \propto {\rm sin}\Psi$ will decrease
as the pitch angle decreases.
We believe this physical effect produces the negative correlation
seen in Figure 4 and dominates over the positive
instrumental correlation between
$\alpha$ and $E_{p}$.
 [The instrumental effect is lessened in this regime because
of the steeper low energy slope (relative to the isotropic case), which allows
the spectrum to reach its low energy asymptote more quickly.]
We point out, however, that the quantitative value of the slope is somewhat shallower
than what is naively expected if {\em only} the mean value of the pitch angle $\Psi$
changes, while all other parameters remain constant.  This is illustrated by the
dashed line in Figure 4, which shows how $E_{p}$ changes as a function
of $\alpha$ 
as the pitch angle $\Psi$ decreases.
To obtain this curve, we simulate SPD spectra for a series of decreasing pitch
angles (given a constant
magnetic field, $\gamma_{m}$, and bulk Lorentz factor $\Gamma$), add Poisson
noise to the spectra,
 and then fit a Band spectrum to each curve; this gives us a value for $E_{p}$ and $\alpha$
 as a function of decreasing pitch angle.  Naturally, dispersion in the magnetic field,
 $\gamma_{m}$ or bulk Lorentz factor 
  will tend to weaken the correlation (as is observed).  We discuss this 
 further in \S 5.1.4 below.
 
\begin{figure}[t]
\plotone{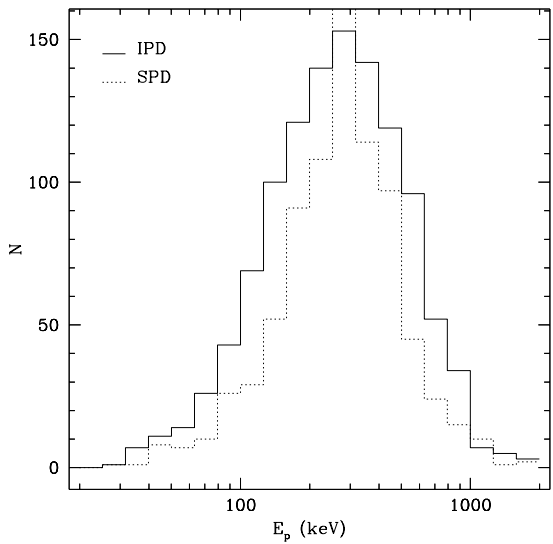}
\caption{The $E_{p}$ distribution for those GRBs in the IPD regime (solid line)
and SPD regime (dotted line).}
\end{figure}

\subsubsection{SAS Regime}
This negative correlation appears to continue for the self-absorbed spectra
(SAS). The low $E_{p}$ value in the SAS regime is consistent
 with
our suggestion (LP00) that the self-absorption frequency is at the
lower edge of the BATSE window (around 50keV). However, there are few bursts 
in this regime and this may be a result
of an observational selection effect.  For very large values of $\alpha$,
bursts with lower values of $E_{p}$ (everything else being equal) will have
a better chance of triggering the BATSE detector.
 Unfortunately, there is not yet sufficient data to test this conjecture.
We note that such a selection effect does {\em not} seem likely to explain
the negative correlation in the SPD regime, because of the 
small difference in the $E_{p}$ distribution between the IPD ($\alpha \la -0.7$)
and SPD ($\alpha \ga -0.7$) cases, as shown in Figure 6.

\subsubsection{Summary}
The $\alpha-E_{p}$ correlations provide, at least qualitatively,
 further support for the synchrotron model and possibly
the small pitch angle distribution (SPD) scenario.  Our most intriguing result
is that at $\alpha \ga -0.7$, the $\alpha-E_{p}$ correlation
goes from positive (as expected purely from instrumental effects discussed above
and in LP00) to negative.  This negative correlation is a natural
expectation in the SPD
 emission regime of our synchrtron models.
However, there are some interesting quantitative points to address.
 The dashed line in Figure 4 gives
an example of how $E_{p}$ should change as a function of $\alpha$ in the
BATSE spectral window,
if only the mean of the electron pitch angle $\Psi$ changes (all
other parameters such as $B$ and $\gamma_{m}$ remaining constant).
The parameter $E_{p}$ - in going from the isotropic to anisotropic
electron distribution regime - should decrease by a factor $\sim \gamma_{m}/2$, as
${\rm sin}\Psi$ decreases from $\sim 1$ to $\sim 1/\gamma_{m}$ (see equations 2 and
3).  It is often assumed that
the electron Lorentz factors in GRBs are $\sim 100$ - approximately equal to 
the bulk Lorentz factors which need to be at least this large 
to keep the medium optically thin to pair production and inverse Compton
scattering (see Lithwick and Sari, 2001, for a recent
discussion of this issue).  In these cases, we should see a decrease
in $E_{p}$ by a factor $\sim 50$ as we transition from the isotropic to
very small pitch angle regime.  Although  this
strong negative correlation will be weakened
to some degree due to dispersion in the intrinsic values of
$\Psi$ and $\gamma_{m}$, as well as
variation in the magnetic field,  
we might expect a stronger decrease in $E_{p}$ 
than what is seen in the data for the large values of $\gamma_{m}$
mentioned above.
The relatively small observed change in $E_{p}$ could mean several things:\\
{\bf (a)} The minimum electron Lorentz factor or the magnetic
field of the electrons
{\em increases} as we transition to a physical regime in which electrons
are accelerated primarily along the magnetic field lines.  This may be
a very plausible explanation - there may exist physical
situations which require either a higher magnetic field
or characteristic electron Lorentz factor, in which it is very efficient to accelerate
along the magnetic field lines.  The details of this 
are beyond the scope of this paper.\\
{\bf (b)} The electron Lorentz factors
could be much smaller than $100$ - in fact, in an internal shocks model, it
is expected that the electron Lorentz factors will be on the order of the
relative Lorentz factor of the two shells, which can be on the order of a
few (Piran, 1999).  This would be consistent with
the relatively small decrease in $E_{p}$ in the SPD regime, seen in the data.
We caution, however, that with electron Lorentz factors so low, we
require a larger value of the $B$ field (for a given observed value
of, e.g., $E_{p}$) and, in addition, some of
the assumptions implicit in our spectral models (e.g. extreme relativistic
velocities) may not hold. \\
{\bf (c)} This model is incorrect and
an alternative explanation is needed to accommodate bursts above the
$\alpha=-2/3$
line of death.  For example,   
the photon indices $\alpha$ between $-2/3$ and
the self-absorption value of $1$ may simply be due to the presence of
{\em both} the absorption frequency and minimum electron frequency present in
the BATSE window.  As long as $\nu_{m} > \nu_{a}$
and spectral fits place the characteristic
break energy at $E_{p} \propto \nu_{m}$, then the low energy photon index will
be a weighted average of $-2/3$ and $1$ (depending on the relative
values of $\nu_{m}$ and $\nu_{a}$). We might then explain the observed
negative correlation between $\alpha$ and $E_{p}$ for $-2/3 \la \alpha
\la 0$ 
by the following: As $\nu_{m}$ decreases (while $\nu_{a}$ remains roughly
constant or even increases), $E_{p}$ decreases and we get less
of the $-2/3$ portion of the spectrum relative to the slope $=1$ portion.  This
will cause the value of
$\alpha$ to increase relative to the $-2/3$ value (of course, in this case, we expect
that the Band fits with a single break will not be as good).  We do not go into further detail
on this subject and only present it as another possibility.

 \subsection{Observed Total photon flux-$E_{p}$ Correlation}
\begin{figure}[t]
\plotone{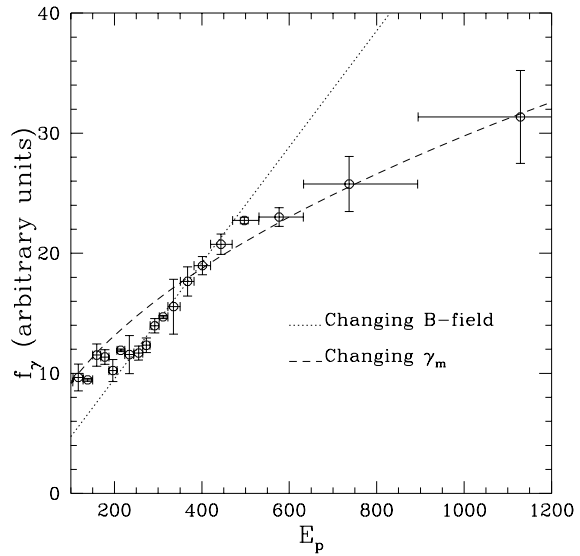}
\caption{ Photon flux $f_{\gamma}$ vs. $E_{p}$ for the binned 2,026
spectra in our sample.  The dotted line shows the correlation (for
all bursts at the same redshift) as a function of an increasing B-field.
The dashed line shows $f_{\gamma}$ vs. $E_{p}$ as $\gamma_{m}$ increases.}  
\end{figure}  
Figure 7 shows a correlation between the average burst total peak flux
and $E_{p}$ (we point out that this correlation is also present
between the normalization or height of the spectrum and $E_{p}$, so
that the value of the peak flux is not greatly affected by the values of the
low and high energy spectral indices and is more a measure of the 
brightness or overall emission power).
The correlation is well established and has also been
reported within individual bursts (Mazets, et al., 2001).
  This correlation can be due to several
effects in a synchrotron model including increasing magnetic
field, minimum electron
Lorentz factor $\gamma_{m}$,
 and/or bulk Lorentz factor $\Gamma$; in Figure 7, we
show
$f_{\gamma}$ as a function of $E_{p}$ in an IPD regime for changing
magnetic field (dotted line) and minimum electron Lorentz factor
(dashed line). We point out again, however,
that the theoretical lines describe the relation between luminosity and
the value of $E_{p}$ in the cosmological rest frame of the burst. A dispersion
in the redshift distribution of the GRBs
will tend to smear out this theoretical correlation. 
We interpret this correlation as a relation between
$E_{p}$ and flux from pulse to pulse (or emission episode 
to emission episode), and have suggested ways to reproduce
this by varying the plasma parameters in the context of
synchrotron radiation.  We note
 that such a correlation has been reported in 
the decay phase of individual {\em pulses} of bursts
 (e.g., Borgonovo and Ryde, 2000, Ryde and Svensson, 2001).
   Such a correlation in a single pulse may arise
from a change in plasma parameters {\em within} a single emission episode, but can also
be produced by relativistic beaming effects alone (photons from the
edge of the beaming cone arrive later than those from the middle of
the cone; the edge photons have a smaller doppler factor [by $1/\Gamma^{2}$]
and so have both a smaller value of
flux {\em and} $E_{p}$ at later times in the pulse). 
 In any case,
the above synchrotron interpretation appears as a plausible candidate for the 
description of the correlation
observed in the data.  

 There may be correlations as a result of other physical
 effects, and correlations between the physical parameters themselves.
 For example, if particle number were conserved in an
 emission episode, then the normalization
 of the particle distribution would increase as $q$ increased - this
 would cause an increase in the overall normalization of the spectrum;
 since the total particle number
 $\int N(\gamma)d\gamma= \int N_{o}\frac{(\gamma/\gamma{m})^{q}}
 {1+(\gamma/\gamma_{m})^{p+q}}d\gamma$, then $N_{o}$ scales
 roughly linearly with $q$.
 
 \section{Examples of Spectral Evolution in Individual Bursts}
 
 Because each pulse is a separate emission episode in our
 model, it is useful to examine what our models can tell us about
 the specific behavior of different internal shocks within an
 individual burst.  This section is intended primarily to
 give the reader a feel for the different types of spectral evolution present in 
 GRBs.  All of the following spectral fits are again taken from
 the catalog of Preece et al. (1999).
 In Figures 8-13, we display the total photon flux, $f_{\gamma}$ (top panels),
 the peak of the $\nu F_{\nu}$ spectrum, $E_{p}$ (middle panels),
 and the low energy photon index, $\alpha$ (lower panels), as a function of time, $t$,
  for several GRBs; the dashed histogram superimposed
 on the $f_{\gamma}(t)$ plot is the time profile (of detector counts) of the burst.
 We now qualitatively discuss the behavior in the context of our synchrotron
 emission scenarios:
 
\begin{figure}[t]
\plotone{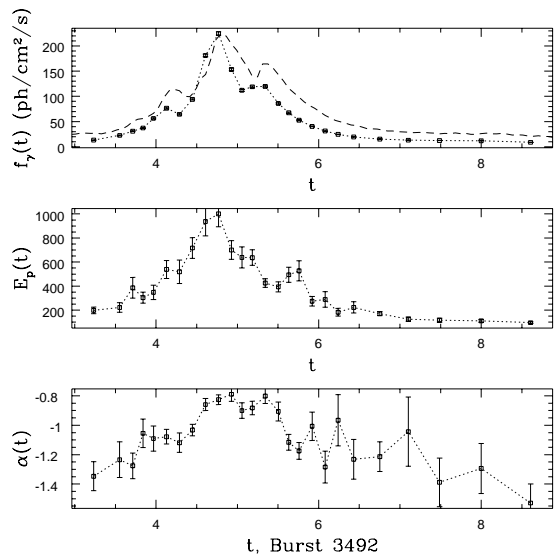} 
\caption{Evolution of photon flux $f_{\gamma}$ (top panel),
$E_{p}$ (middle panel), and $\alpha$ (lower panel) for burst 3492. The dashed line
in the top panel is the total GRB count rate arbitrarily normalized to the peak
flux value.  Note that this burst remains
in the IPD regime throughout.}
\end{figure}  

\begin{figure}[t]
\plotone{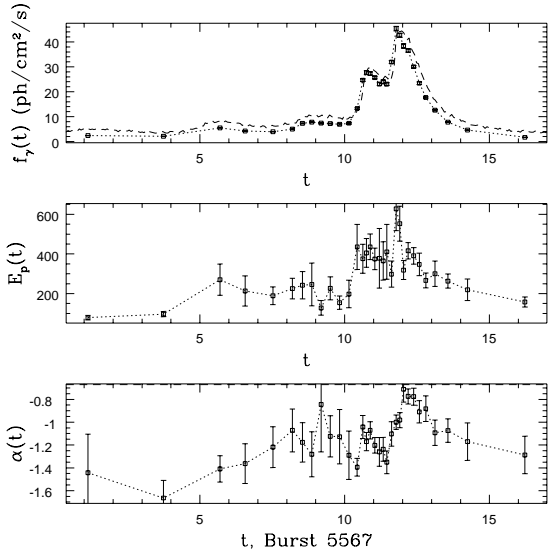}
\caption{ Same as Figure 8, but for
burst 5567.}
\end{figure}
  
{\bf 1. Burst 3492:} Figure 8 displays the spectral
 evolution of burst BATSE trigger number 3492.  This is a good example
 of how $E_{p}$ tracks the flux (or counts) of the burst.  The parameter
 $\alpha$ somewhat tracks $E_{p}$ in this case, but there is evidence for
 a change in this trend which could be indicative
 of a change in the lower cutoff parameter $q$ of the electron spectrum.
 We note that - according to its value of $\alpha$ - this burst remains
 in the IPD regime throughout its duration.  
 
 {\bf 2. Burst 5567:} Figure 9  is another example of $E_{p}$ tracking
 the flux and $\alpha$ tracking $E_{p}$ - entirely in the IPD regime.
 
 {\bf 3. Burst 2286:}  In Figure 10, $E_{p}$ appears to correlate with the flux,
 and begins at low or soft values and ends at higher, harder values (note
 that the first point in the plot at $t=0$ could be indicative of
 a precursor).
This suggests that during the later emission episodes,
 either the magnetic field or minimum Lorentz
 factor changed to cause an increase in $E_{p}$.  How this is physically
 achieved is a complicated issue, and is related to the level of turbulence
and/or the specifics of the particle acceleration mechansim in the shock.
  The interpretation of spectral evolution in our models gives us a way
  of directly interpreting how the physics can change from shock episode
  to shock episode.
   
 {\bf 4. Burst 2855:} In Figure 11, we see that $\alpha$ evolves
 from about $1$, interpreted as
 a self-absorbed situation (and note the low value of $E_{p}$), to about
 zero, a value appropriate for the  small
 pitch angle case, and then back to $1$ or
 the self-absorbed regime at the end of the burst (where
 again $E_{p}$ is at its lowest).  If this interpretation is correct, the physics 
 required to produce this behavior is intriguing to say the least - particularly
 switching back to the self-absorbed regime at the end of the burst. 
 We also point out that for this burst - which is
 entirely in the SPD and SAS emission regimes - $\alpha$ appears to roughly
 anti-correlate with $E_{p}$ (see \S 3.1 and Figure 4 above; although admittedly,
 the error bars are large here and a constant $E_{p}$ is a statistically
 acceptable - although not the best - description of the data).  
 
 {\bf 4. Burst 3489:} Figure 12 shows an additional example in which 
$\alpha$ appears to roughly anti-correlate with $E_{p}$ during the various
emission episodes of this burst, particularly from $t\sim 9 \rm s$ to $14 \rm s$.
 
  {\bf 5. Burst 1886:}  Figure 13 shows an example of
 $\alpha$ transitioning from the SAS to SPD  regime; meanwhile, $E_{p}$
 appears to roughly track the flux of the burst. 
 
\begin{figure}[t]
\plotone{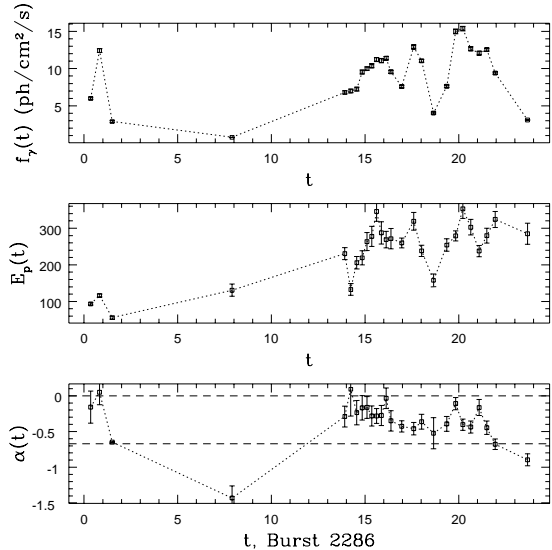}
\caption{Same as Figure 8, but for
burst 2286.  In the bottom panel, the upper and lower dashed horizontal lines delineate
the SPD and IPD regimes, respectively.}
\end{figure}

\begin{figure}[h]
\plotone{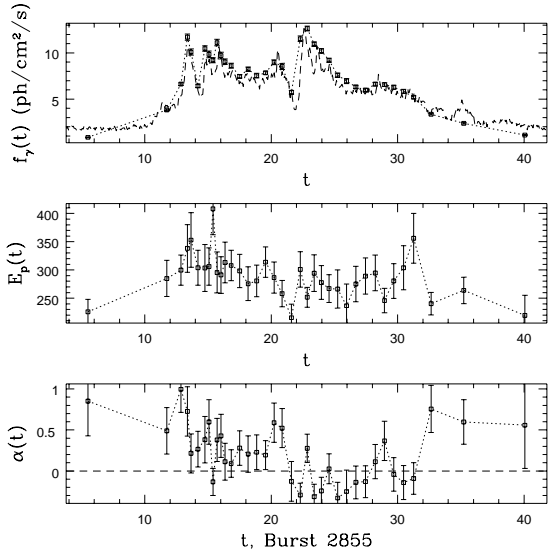}
\caption{Same as Figure 8, for burst 2855. The horizontal dashed line separates the SAS (above
the line) and SPD (below the line) regimes. Note that this burst remains entirely in
the SPD and SAS regimes througout its entire duration.  There is also a rough anti-correlation
between $\alpha$ and $E_{p}$ for this burst.}
\end{figure} 

\begin{figure}[h]
\plotone{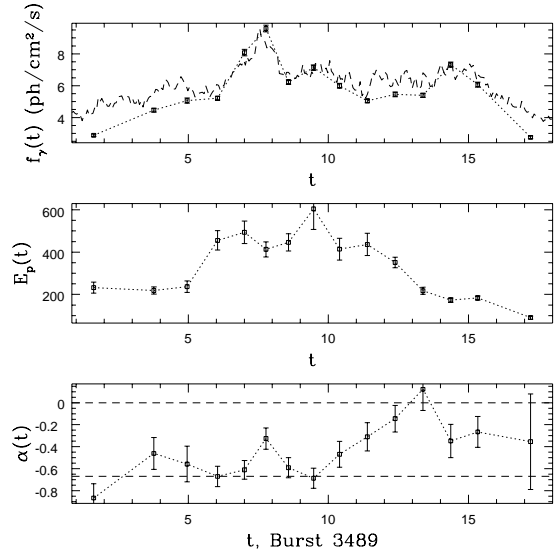}
\caption{Same as Figure 8, but for
burst 3489.  During the various emission episodes (to the extent they
can be delineated), $E_{p}$ is on average lower for the higher values
of $\alpha$ in the SPD regime.}
\end{figure} 

\begin{figure}[h]
\plotone{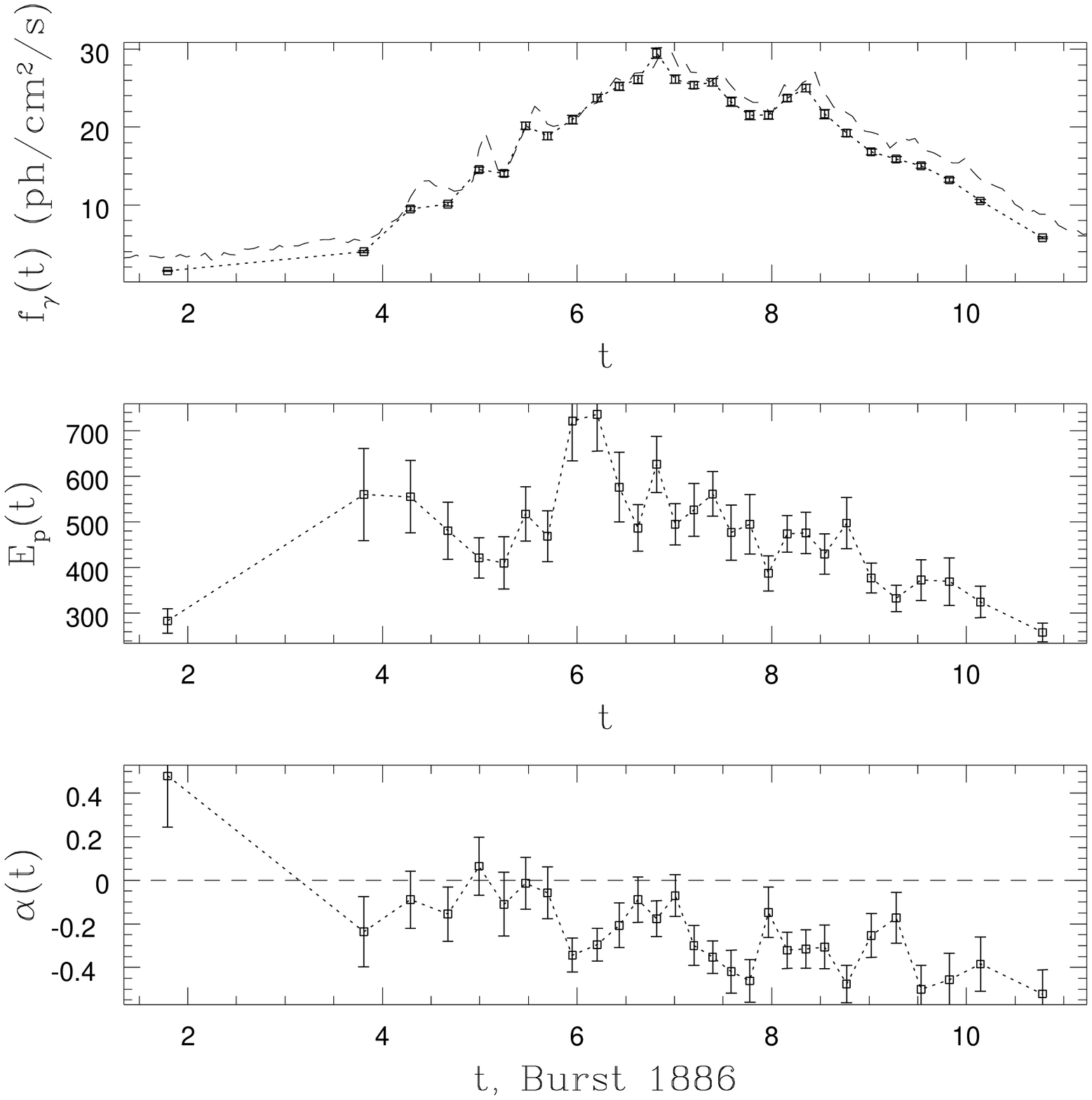}
\caption{Same as Figure 8, but for
burst 1886.  The horizontal dashed line delineates the SPD  emission regime.
This burst evolves from an SAS to SPD regime.}
\end{figure}

 The different types of
  behaviors seen in the individual bursts
  are interesting. On one hand, the spectral evolution
   appears to be highly varied and all
  types of behavior can exist (see also Kargatis, 1994). On
  the other hand, certain trends appear to be present in the data
  that are at least qualitatively
  consistent with the global trends presented in \S 4. Of course, we acknowledge
  that these global trends will not be seen in all bursts; establishing statistically
  robust trends for each particular emission regime in each particular burst in
  an attempt to say something about the {\em average} trends among all bursts
  is not statistically meaningful (see section 4 for discussion of this issue).
  We do emphasize again that all types of behavior can occur within individual
  bursts, and we can use our models to interpret the physics behind any
  particular evolution. For example, Crider et
  al. (1997) find that in GRB910927\footnote{Note that
  this burst is not present in the Preece et al. catalog
  because it does not meet one of the catalog's selection criteria - 
  namely, a peak flux greater than $10$ ph/cm$^2$/s on the
  1024ms timescale.}, there is a positive correlation
  between $\alpha$ and $E_{p}$ in the  SAS regime (and very slightly in the SPD
  regime, although there are only 6 points here).  This burst appears
  to have one (or at most, two) smooth, broad pulses, so that there is
  probably only one (or two) emission episode(s) giving rise to the various spectra.
  A positive correlation in this case, could be due to a number
  of factors in our model. Of course, as discussed in LP00 and
  in \S 2, the finite bandwidth of the BATSE spectral window will tend
  to produce a positive correlation between $\alpha$ and $E_{p}$. However,
  a physcial explanation is also plausible; for example, it is possible
  that the the self-absorbed (or small pitch angle) portion of the
  spectrum is decreasing relative to the IPD ($\alpha=-2/3$) portion
  due to a decrease in, say, the magnetic field in the shock. If
  $E_{p}$ corresponded to the self-absorption frequency,
   this would produce a positive correlation
  between $E_{p}$ and $\alpha$.  This correlation would be exacerbated if
  the cutoff to the low end of the electron distribution evolved to a more flat
  distribution (lower $q$ values) throughout the pulse.
  
   Our point is that no matter what
  the behavior, our models can be used as diagnostics
  to interpret what types of physical changes may be occuring
  from pulse to pulse.   
  Note that if we can in fact characterize our
 synchrotron emission regime by the value of $\alpha$, then it is particularly
 interesting when we see bursts switch emission regimes from pulse to pulse.
 (as a caveat, however, we note that the error bars are significant
  on some of the spectral fits and a clear cut interpretation
  is not always readily available).  
 This suggests that the fundamental plasma physics (i.e. the particle
 acceleration) can vary depending on the internal
 shock conditions within a single
 burst.   
 
\section{Discussion and Conclusions}
In this paper, we have explored the validity of the synchrotron
model in explaining the
behavior of the time resolved GRB spectra.
In our model, there are three different emission regimes, all distinguished
by the value of the low energy photon index $\alpha$.  Our model
also accounts for the instrumental correlation between $\alpha$ and
$E_{p}$ produced when $E_{p}$ is near the lower end of the BATSE
spectral window (in this case, the spectrum has not reached its low
energy asymptotic value and so $\alpha$ will be {\em softer}
or lower for smaller $E_{p}$'s).  In the IPD
 regime, the synchrotron emission is optically
thin from an isotropic electron distribution, and the low
energy photon index $\alpha$ is less than or equal to $-2/3$.  In
the SPD regime, the emission is primarily from electrons with small
pitch angles and the spectrum consequently has a value of $\alpha \sim 0$.
Finally, in the SAS regime, the synchrotron photons are self-absorbed and
there is a steep cutoff ($\alpha = 1$ or $3/2$ depending on the relative
values of the absorption and minimum electron frequency) in the low energy
spectrum.  

We have shown that these models provide an excellent description of the
existing data, and have presented spectral fits for a few bursts.  
Our results suggest that the Band parameter $\alpha$ is a good diagnostic
of the relevant emission regime.  We also have presented the spectral evolution
of the Band spectral parameters (specifically photon flux, $E_{p}$, and
$\alpha$), and interpret the behaviors in the context of our model.  We point
out that the behavior is quite varied and there is often evidence that a burst
switches emission regimes from pulse to pulse.  In attempt to characterize
general evolutionary trends exhibited in GRBs, 
we have combined all 2,026 separate
time resolved spectra (from the 80 bursts in our sample) and looked for
correlations 
in this data.  Our main results are as follows:\\ 
$\bullet$ We find that the majority of the bursts lie
in the IPD regime ($\sim 55\%$), but that a
significant fraction  of the bursts ($\sim 40 \%$) are in the SPD regime.
This has interesting implications for the particle acceleration studies
in GRBs as we discuss below.
 We find that
only a small fraction ($\sim 5\%$) of bursts are in the self-absorbed (SAS)
regime.\\
$\bullet$ We show that there is a strong correlation between $\alpha$
and $E_{p}$ in the IPD regime, which can be attributed to the 
positive instrumental correlation discussed in LP00. 
However, there appears to be a negative correlation
between $\alpha$ and $E_{p}$ in the SPD regime (albeit with $4\sigma$
significance).  We interpret this
as evidence of the effects of decreasing electron pitch angle as
$\alpha$ transitions from $-2/3$ to $0$.  This physical effect overwhelms
the positive instrumental correlation that is dominant in the IPD regime.
However, for this physical effect to accommodate the data quantitatively,
we need either an increase in the minimum electron Lorentz factor as
we transition to the SPD regime, or for the absolute value of the
electron Lorentz factor to be only on the order of a few. This may very
well be the case in an internal shocks model, where the {\em relative}
Lorentz factor of the two shocks is the relevant scale for the electron
Lorentz factors.\\
$\bullet$  We find a strong positive correlation between the photon flux and
$E_{p}$, which can be explained by several effects  in a synchrotron emission
model.  Most notably, a change in the magnetic field or minimum Lorentz factor
will produce this type of a correlation.\\

Our results bring to light the fact that particle acceleration in GRBs
is a quite poorly understood problem.  Usually, it is assumed
that the radiating particles in GRBs are accelerated via repeated
scatterings across the (internal) shocks.
This is because shocks 
can quickly accelerate particles to very high energies,
through repeated scatterings across the shock
(the scattering agent being
plasma turbulence from, e.g., a two-stream instability Medvedev and
Loeb, 1999).
  This mechanism, however,
predicts several features in the electron distribution not borne out
by the data.  First, it 
 has been shown (Kirk et al., 2000 and references therein)
that these
repeated crossings of the shock result 
in a power law particle
distribution with a well defined index, $p = -2.23$, which would give a 
high energy synchrotron photon index $\beta$ of -1.62 (or -2.12 for the
``cooling'' spectrum, e.g. Sari et al., 1997).
 Although this is consistent with some afterglows, this is
certainly is not true for many bursts in the 
prompt phase.
In our synchrotron models above, the high energy photon index 
$\beta=-(p+1)/2$, where $p$ is the high energy index of the emitting
particle distribution.  The parameter
$\beta$ can vary by a factor of
4 (or more!) throughout a single burst (see, e.g., Preece et al., 1999),
 reflecting a huge variation (from $1$ to $9$) in the parameter
$p$ of the underlying
particle distribution - this is well beyond the statistical limits placed on
$p$ by shock acceleration simulations.
In addition, shock acceleration predicts an {\em isotropic}
distribution of electrons.  Our work 
 suggests that in a large fraction of GRBs, the
particle  acceleration is not isotropic but along the magnetic field lines.

In fact, when the
 Alfv\'en (phase) velocity
$\beta_A=B/(4\pi 
nm_pc^2)^{1/2}$ (in units of the speed of light)
is greater than $1$, stochastic acceleration becomes
more efficient than shock acceleration (Dung and Petrosian, 1994).
It is well known that in this case the electric 
field fluctuations $\delta E \sim \beta_A \delta B$ exceed the magnetic field 
fluctuations, which means a faster acceleration ($\propto \delta E^2$) 
than scattering rate ($\propto \delta B^2$). 
Under these 
conditions, once the particle crosses the shock front into the turbulent region 
behind the shock, it will undergo stochastic acceleration much faster than it 
can 
be turned around to cross the shock again.  This is the
situation for GRBs which have inferred magnetic fields of $B \sim 10^{5} G$, and
densities of $n \le 10^{8} cm^{-3}$ so that $\beta_{A} \gg 1$. 
  The shape of the spectrum is then determined by the relative values of the 
diffusion coefficients and the rates of energy and pitch angle changes resulting
from the interaction of the injected particles with the plasma turbulence behind the shock.
 Hence, we emphasize the importance of an
  investigation of particle acceleration
 in GRB internal shocks.  
 
  Lastly, we would like to emphasize the important role upcoming (and
  current) GRB missions will have in laying to rest the questions raised
  in our investigation.  With the launch of HETE-2, and the upcoming
  launches of Swift and GLAST,  we will be able to get high quality broadband  
  (from a few eV to GeV) spectra of the prompt GRB
  emission; such spectra will allow us to test our models  more
  stringently and constrain all of the emission mechanisms that may
  play a role in GRBs (for example, we may see a synchrotron self-Compton
  component in the GLAST energy range; Dermer et al., 2000).  Data from these satellites 
  are sure to shed significant light on the photon spectrum and particle
  acceleration mechanisms in Gamma-Ray Bursts.  
  
  Acknowledgments: We thank the anonymous referee for a careful
  report which led to improvements on this paper. We
  also thank Felix Ryde for interesting discussions.
  We acknowledge funding by NASA grant NAG5-7144.

 \end{document}